\documentstyle[psfig]{elsart}

\newcommand{\be}{\begin{equation}}
\newcommand{\ee}{\end{equation}}
\newcommand{\bea}{\begin{eqnarray}}
\newcommand{\eea}{\end{eqnarray}}
\begin{document}
\begin{frontmatter}


\title{Excess low energy photon pairs from pion annihilation
 at the chiral phase transition}

\author[dubna]{M.K.~Volkov\thanksref{intas1},}
\author[dubna]{E.A.~Kuraev\thanksref{intas2},}
\author[rostock]{D.~Blaschke,}
\author[dubna,rostock]{G.~R\"opke,}
\author[rostock,telaviv]{S.~Schmidt\thanksref{minerva}}

\address[dubna]{Bogoliubov Laboratory for Theoretical Physics \\
        Joint Institute for Nuclear Research, \\
        14 19 80 Dubna, Russia  }
\address[rostock]{Fachbereich Physik, Universit\"at Rostock\\
	D-18051 Rostock, Germany}
\address[telaviv]{School of Physics and Astronomy \\
	Raymond and Beverly Sackler Faculty of Exact Sciences \\
        Tel Aviv University, 69978 Tel Aviv, Israel}

\thanks[intas1]{Supported by INTAS grant No. W 94-2915}
\thanks[intas2]{Supported by INTAS grant No. W 93-0239}
\thanks[minerva]{Supported by MINERVA Foundation}

\begin{abstract}
The photon pair production by pion annihilation in a hot and dense medium at 
the chiral phase transition is investigated within a chiral quark model.
As a direct consequence of this transition the $\sigma$ meson appears as a 
bound state in the domain of temperatures and chemical potentials where the 
condition $M_\sigma(T,\mu) \approx 2\,M_\pi(T,\mu)$ is fulfilled. 
This effect results in a strong enhancement of the cross section  for
the pion annihilation process $2 \pi \rightarrow 2 \gamma$
compared with the vacuum case. The calculation of the photon pair production 
rate as function of the invariant mass shows a strong enhancement and narrowing
of the  $\sigma$ meson resonance at threshold due to chiral symmetry 
restoration.

\vspace{5mm}

\noindent 
PACS number(s):  05.70.Jk, 11.30.Rd, 13.40.-f, 14.40.-n, 25.75.-q
\end{abstract}
\end{frontmatter}

\newpage

\section{Introduction} \label{introsec}

The investigation of hadron properties at finite temperature and density is
very important in order to provide insight into phenomena occuring in  the
vicinity of the transition to the hypothetical quark-gluon plasma state of
matter which is characterized by hadron deconfinement and restoration of
chiral symmetry.
These studies become particularly interesting since in ultra-relativistic
heavy ion collision experiments performed and planned at BNL Brookhaven and
CERN Geneva one is able to create matter at the extreme densities and
temperatures necessary for the phase transition.
Different effects which were suggested as possible signatures for the
chiral/deconfinement transition such as $J/\Psi$ suppression, strangeness
enhancement and low-mass dilepton enhancement have been observed.
For a recent discussion of these experimental findings, see \cite{qm95}.
The most direct probes from the suspected quark-gluon plasma state and the 
subsequent hadronisation transition are photons and dileptons since they leave 
the hot and dense matter at an early stage
without suffering strong interactions.

The most abundant secondaries produced in heavy ion collisions are pions which
evolve through a hot and dense hadron gas state until the freeze-out.
Therefore it is interesting to investigate the behaviour of such a gas at
large temperature and/or density, in particular to study its radiation. 
The main channel of this radiation is the annihilation of two pions into a 
virtual photon which decays into a lepton pair. 
With help of this reaction it is possible to study the behaviour of 
the intermediate $\rho$ meson under the influence of a hot and dense medium 
\cite{rapp,brown,hajo}. 

Besides of this dominant radiation process of the pion gas, 
the annihilation of two pions into two photons is of particular importance 
since its cross section is
sensitive to changes in the $\sigma$ meson properties which occur in the 
vicinity of the chiral restoration transition.
In the vacuum this reaction contributes to the  amplitude
with noticeable probability only via the Born diagrams for the charged pions.
Compared with the vacuum case, with increasing temperatures and densities the 
$\sigma$ meson \cite{4,6,1,osip} changes its character \cite{HK} from a 
broad resonance with a large decay width into the two-pion channel
to a bound state below the two-pion threshold 
$M_\sigma(T,\mu) \le 2 M_\pi(T,\mu)$ and 
its contribution to the cross section of the process $2\pi\to2\gamma$ 
overwhelms that of the Born terms \cite{preprint}.
This effect can be understood as an inevitable consequence of the transition
from a phase of matter with restored chiral symmetry (i.e. degenerate 
scalar and pseudoscalar fields) to a phase where this symmetry is broken and 
the scalar meson aquires a mass above the two-pion threshold with a large decay
width into the two-pion channel. 
A similar effect has been discussed by Weldon \cite{weldon} for the soft
lepton pair production by pion annihilation in a hot, dense pion gas where due
to chiral symmetry restoration the $\sigma$ resonance can give a sizeable 
contribution to lepton pair spectra provided a charge asymmetry of the pionic 
medium.

It is the aim of the present paper to investigate the enhancement of 
low-energy photon pairs from pion annihilation at the chiral/hadronisation
transition and to give numerical estimates for both the magnitude of the cross 
section enhancement (critical scattering) and the photon production rate on 
the basis of a chiral quark model.
This effect of a threshold enhancement in the photon pair spectra 
could indicate chiral symmetry restoration from
ultrarelativistic heavy-ion collisions.
 
The decay of the pion into two quarks appears at higher values of $T$ and 
$\mu$ and does not play any r\^ole in the chosen scenario. But it is important 
to note that in this domain the $\sigma$ meson enters the continuum of 
$q{\bar q}$ states. The appearence of such a $\sigma$ resonance and the 
photon pair spectrum in a quark plasma has been studied in \cite{reh} and 
complements the present investigations.

\section{The process $\pi \pi \rightarrow\gamma\gamma$ at finite temperature 
and chemical potential}
\label{sec:pipi}

The process $\pi \pi \rightarrow\gamma\gamma$ is described by the Born terms
(for charged pions, see Fig. 1) 
and the quark-substructure terms (for charged and neutral pions, see 
Fig. 2)\footnote{The Fig. 1 is the first (local) approximation for the 
diagrams of Fig. 2 a,b}. 
The box diagrams and the $\sigma$ pole diagram (Fig. 2 b, c) are the most
important contributions.
Processes with vector mesons ($\rho, \omega$) in the t-channel are also
present but give a negligible contribution.

The Lagrangian describing the diagrams of Fig. 1 has the form
\bea
\label{L1}
{\cal L} = ie A_\mu(\pi ^-\partial_\mu\pi ^+-\pi ^+\partial_\mu\pi ^-)
+ e^2{\cal A}_\mu^2~\pi ^+\pi ^- ~ .
\eea
The diagrams of Fig. 2 describe the most important contributions to the 
$\pi \pi \rightarrow\gamma\gamma$ process in terms of a low energy quark model
of QCD in $1/N_c$ approximation. 
According to this level of description the meson-meson and meson-photon 
vertices are given by quark-loop diagrams (triangle and box diagrams) which 
have previously been 
analysed, e.g., within the Nambu--Jona-Lasinio (NJL) model \cite{1,osip,2}.
In local approximation and for low-energy photons ($k^2$ approximation) the 
relevant vertices are given by the following Langrangians 
\bea
\label{Lbox}
{\cal L}_{{\rm box}} &=& \frac{\alpha}{18\pi  f^2_\pi }
 (\pi^+\pi^- + 5\pi^0\pi^0)F^2_{\mu\nu}\,,\\
{\cal L}_{\sigma\pi\pi}&=&2mg_\pi\sigma(2\pi^+\pi^- +\pi^0\pi^0),\\
{\cal L}_{\sigma\gamma\gamma}&=& \frac{5 \alpha}{9\pi f_\pi }
\sigma F^2_{\mu\nu},
\label{L2}
\eea
where
$\alpha =e^2/4\pi =1/137$,
$F_{\mu\nu}=\partial_\mu A_\nu-\partial_\nu A_\mu$.
The components of the isovector pseudoscalar meson field are denoted by 
$\pi^+,\pi^-,\pi^0$; $\sigma$ is the isoscalar, scalar meson field.
The values of the dynamical quark mass $m$, the pion decay constant
$f_\pi$,  the strong pion coupling constant
$g_\pi = m/f_\pi$ as well as the meson masses $M_\pi$ and $M_\sigma$ 
at finite temperatures $T$ and chemical potentials $\mu$ are taken from Ref.
\cite{sbk,doc} for the case of the NJL model \cite{njl}.

The amplitude of the process $\pi^+\pi^- \to 2 \gamma$ 
describing the sum of the Born diagrams as
well as the $\sigma$ pole and the box graphs is given by
\bea
\label{amplitude}
T_{\pi^+\pi^-\rightarrow\gamma\gamma} = e^2\bigg[ 2\bigg(g^{\mu\nu}-
\frac{q_1^\mu q_2^\nu}{q_1k_1}-\frac{q_2^\mu q_1^\nu}{q_2k_1}\bigg)+{\cal
A}_{\pi^+\pi^-\rightarrow\gamma\gamma}
(g_{\mu\nu}k_1k_2 - k_1^\nu k_2^\mu)\bigg ]e_\mu(k_1)e_\nu(k_2),
\eea
where $q_i$ and $k_i$ are the momenta of pions and photons, respectively,
$e_\mu(k_1)$ and $e_\nu(k_2)$ denote the polarization vectors of the
photons. The amplitude
\bea
\label{amp}
{\cal A}_{\pi^+\pi^-\rightarrow\gamma\gamma} = \frac{1}{(6\pi f_\pi (T, \mu)
)^2}\bigg[ \frac{40m^2(T, \mu)}{M^2_\sigma(T, \mu)-s-iM_\sigma(T, \mu)
\Gamma_\sigma(T, \mu)}f_1(\mu,T) - f_2(\mu,T)\bigg]
\eea
includes the quark substructure contributions of the $\sigma$ pole and the 
box diagrams of Fig. 2. 
We refer the reader to Refs. \cite{1,osip,2} for details of the derivation 
of Eqs. (\ref{amplitude}), (\ref{amp}).
Mass and decay width of the $\sigma$ meson are given within this model as
\bea
\label{msig}
M^2_\sigma (T,\mu)&=& M^2_\pi(T,\mu) + 4m^2(T,\mu)~,\\
\label{gsig}
\Gamma_\sigma (T,\mu)&=& \frac{3[M^2_\sigma (T,\mu)- M^2_\pi(T,\mu)]^2}
{32\pi M_\sigma(T,\mu) f^2_\pi(T,\mu) }
\sqrt{1 -\frac{4M^2_\pi(T, \mu) }{M^2_\sigma(T, \mu)}}~.
\eea
The mass of the $\sigma$ meson is determined by its internal quark 
substructure and calculated by solving the Bethe-Salpeter equation at finite 
temperature $T$ and chemical potential $\mu$ in the scalar quark-antiquark 
channel. In the NJL model this leads unambiguously to relation (\ref{msig}).
The width of the $\sigma$ meson (\ref{gsig}) is given by the strong 
decay channel into two pions which is evaluated in the standard way, e.g. 
within the linear sigma-model, see e.g. \cite{HK,weldon}. 
The $T, \mu$ dependences of $M_\sigma, M_\pi$ and
$f_\pi$ are obtained from the NJL model.  
The functions $f_1(T, \mu)$ and $f_2(T, \mu)$ describe the dependence of the
quark triangle and the quark box diagrams on temperature and chemical 
potential \cite{2,mkv} and are given by
\bea
\label{f1}
f_1(T,\mu)&=&1-\frac{3}{2}m^2(T, \mu)\int_0^\infty dk \frac{k^3}{E^6(k)}
\ln\bigg[\frac{E(k)+k}{E(k)-k}\bigg]
\big[ n(k;T,\mu)+{\bar n}(k;T,\mu)\big]~,\\
\label{f2}
f_2(T,\mu)&=&3m^2 (T, \mu)\int_0^\infty dk\frac{k^2}{E^5(k)}
\big[1- n(k;T,\mu)-{\bar n}(k;T,\mu)\big]~.
\eea
The Fermi distribution functions
$n(k;\mu,T)=\{1+\exp[(E(k)-\mu)/T]\}^{-1}$ for particles and
${\bar n}(k;\mu,T)=\{1+\exp[(E(k)+\mu)/T]\}^{-1}$ for antiparticles
describe the dependence on temperature and chemical potential for quarks
and antiquarks of the energy $E(k) = \sqrt{k^2+m^2(T, \mu)}$. 
Note that the occurence of the functions 
$f_{1,2}(T,\mu)$ is an important point of our study: 
the local vertices (Fig. 1) of the Born approximation  are replaced by quark
loop diagrams (Fig. 2 a,b). Therefore the vertices become non-local and suffer 
an explicit medium dependence due to the quark substructure.

In the following we use a short notation, where we drop the
$T$- and $\mu$ dependences in the arguments of the medium dependent
quantities.
For the square of the transition matrix element (\ref{amplitude}) we obtain
\small
\bea
|T_{\pi^+\pi^-\rightarrow\gamma\gamma}(s)|^2 &=& 4e^4\bigg (2+M^4_\pi \bigg
[\frac{1}{(k_1q_1)^2}+
\frac{1}{(k_2q_1)^2}\bigg ]
+ 2\frac{(q_1q_2)^2}{(k_1q_1)(k_2q_1)}-2q_1q_2(\frac{1}{k_1q_1}
+\frac{1}{k_1q_2})\nonumber\\
&+&{\rm Re}[{\cal A}_{\pi^+\pi^-\rightarrow\gamma\gamma}(s)]\bigg
[3k_1k_2+\frac{(k_2q_1)^2-(k_1k_2)(q_1q_2)}{k_1q_1}
+\frac{(k_1q_1)^2-(k_1k_2)(q_1q_2)}{k_2q_1} \bigg]\nonumber\\
&+&\frac{|{\cal A}_{\pi^+\pi^-\rightarrow\gamma\gamma}(s)|^2}{2}(k_1k_2)^2
\bigg)\,.
\eea
\normalsize
With the following definitions of the variables:
$k_1=(\omega, {\bf k})$,
$k_2=(\omega, -{\bf k})$,
$q_1=(\omega, {\bf q})$,
$q_2=(\omega, -{\bf q})$,
$s = (q_1+q_2)^2=4\omega^2$,
the calculation of the cross section for this process gives
\bea
\sigma^{\pi ^+\pi ^-\rightarrow\gamma\gamma} (s)&=&
\frac{1}{16 \pi^2}
\frac{1}{s\kappa}\int\frac{d^3k_1}{2\omega_1}\frac{d^3k_2}{2\omega_2}
\delta^4(q_1+q_2-k_1-k_2)|T_{\pi^+\pi^-\rightarrow\gamma\gamma}(s)|^2
\nonumber\\
\label{spiga}
&=&\sigma_1(s)+\sigma_2(s)+\sigma_3(s)~,
\eea
with the partial cross sections
\bea\label{sig1}
\sigma_1(s)&=&16\sigma_0~\bigg(2-\kappa^2 -\frac{1-\kappa^4}{2\kappa}
\ln\bigg[\frac{1+\kappa}{1-\kappa}\bigg]\bigg),\\\label{sig2}
\sigma_2(s)&=&4\sigma_0~ s~ {\rm Re}\bigg[{\cal
A}_{\pi^+\pi^-\rightarrow\gamma\gamma}(s)\bigg]\frac{1-\kappa^2}{\kappa}
\ln\bigg[\frac{1+\kappa}{1-\kappa}\bigg],\\
\sigma_3(s)&=&\sigma_0~s^2~ |{\cal A}_{\pi^+\pi^-\rightarrow\gamma\gamma}(s)|^2
\label{sig3}
\eea
corresponding to the Born, the interference and the quark substructure
contributions, respectively, see Figs. 1 and 2. 
We have introduced the abbreviations 
$\sigma_0=\pi\alpha^2/4s\kappa$ and $\kappa^2=1-4M_\pi^2/s$.

In order to give estimates if an enhancement of the cross sections causes an
enhancement of the photon pair production rate, we calculate 
\bea\label{rate}
\frac{dN_{\gamma\gamma}}{d^4xdM}=4M\int\frac{d^3p_1}{(2\pi)^3}
\int\frac{d^3p_2}{(2\pi)^3}v_{\rm rel}
\sigma^{\pi^+\pi^-\rightarrow \gamma\gamma}
n_\pi(p_1)n_\pi(p_2)\delta\big(M^2-(p_1+p_2)^2\big)
\eea  
being the rate per space-time element with the relative velocity 
$v_{\rm rel}(p_1,p_2)=\sqrt{1-M_\pi^4/(p_1p_2)^2}$ and the pion distribution 
function $n_\pi(p)=\{\exp(\sqrt{p^2+M_\pi^2}/T)-1\}^{-1}$. 
After integration over momenta and angles one obtains
\bea
\frac{dN_{\gamma\gamma}}{d^4xdM}=&&\frac{M^2T^2}{(2\pi)^4}
\sigma^{\pi^+\pi^-\rightarrow\gamma\gamma}\sqrt{M^2-4M_\pi^2}
\int^\infty_\lambda dx\frac{1}{e^x-1}\ln\bigg[
\frac{1-\exp(F_-(x,T,\mu))}{1-\exp(F_+(x,T,\mu))}
\bigg]\,,
\eea
where $\lambda = M_\pi/T$ and
\[
F_\pm(x,T,\mu) = -x\bigg(\frac{M^2}{2M_\pi^2(T,\mu)}-1\bigg)\pm\sqrt{x^2-\lambda^2}\sqrt{\bigg[\frac{M^2}{2M_\pi^2(T,\mu)}-1\bigg]^2-1}\,.
\]
Similar expressions can be found in \cite{kare} for dilepton production rates. 
Note that the cross section 
$\sigma^{\pi^+\pi^-\rightarrow\gamma\gamma} = 
\sigma^{\pi^+\pi^-\rightarrow\gamma\gamma} (s = M^2,T,\mu)$ 
is medium dependent and obtained from (\ref{spiga}). 

In the following, we examine the change of the cross section for photon pair 
production and the photon production rate from  two-pion annihilation in a hot and dense medium numerically.
To this end, we use as an input the in-medium modifications of quark and meson
properties which we have obtained from a chiral quark model at finite 
temperature and chemical potential.

\section{Numerical results}

In our model we fix the constituent quark mass\footnote{We have used a 
3-dimensional cut off for regularization of the divergent integrals.} in the 
vacuum at $m = 300$ MeV 
and obtain at zero temperature and chemical potential the values for the meson 
masses and the pion decay constant given in Table 1. 
At finite temperature and chemical potential these values change due to medium 
effects what has been analyzed in  several publications, e.g. \cite{sbk,doc,5}.
In the following we study the behaviour of the cross sections at special points
in the $T-\mu$ plane. The calculation of the quark and meson properties within
the NJL model  provides the values shown in Table 1 for $T=100$ MeV and 
$\mu= 250$ MeV and $T=170$ MeV and $\mu=80$ MeV. The choice of the latter set 
is motivated by a recent analysis of hadron spectra from heavy-ion collisions 
at SPS energies within a thermal model \cite{sps}. 
From the finite temperature studies of the chiral phase transition we know that
due to the chiral symmetry restoration there occurs also an ``inverse'' Mott 
transition where the $\sigma$ meson mass drops below the two-pion continuum
threshold and thus the width for the $\sigma \to \pi\pi$ decay vanishes. 
Consequently the $\sigma$ pole contribution to cross section becomes 
enhanced close to the two-pion threshold. This effect has been paraphrased as
``critical scattering'' in the analysis of cross sections for hadronization 
\cite{huefner} and two-photon production in a quark-meson plasma \cite{reh}. 

We have calculated the cross section for the annihilation of two charged 
pions into two photons and we have analyzed the temperature and density 
dependence of the different contributions (\ref{sig1}-\ref{sig3}) to the 
total cross section (\ref{spiga}). 
Due to  substructure effects we expect to obtain a strong enhancement of the 
total cross section in vicinity of the two-pion threshold. 
In Fig. 3 we show as our main result that the cross section for two-photon
production is critically enhanced only in a narrow band of values in the 
$T-\mu$ plane which is closely related to the critical line for the 
chiral restoration transition.

Let us  study this effect more in detail. 
The different partial contributions to the cross section as a function of the 
energy $s$ are given in Fig. 4. The upper panel shows the numerical 
results for the vacuum ($T=\mu=0$) case where the main contribution comes from 
the Born diagram while the substructure has only a small contribution of about 
$10\%$. 
The situation changes at finite temperature and chemical potential. 
While the Born term does not change, the substructure terms become important. 
As shown in the lower panel of Fig. 4, the $\sigma$ pole diagram gives the 
dominant contribution and the total
cross section is strongly enhanced at threshold.

For the neutral pion annihilation we obtain qualitatively the same  result. 
In this case the Born diagrams do not contribute at all and we have a drastic
increase of the cross section at the threshold by three orders of magnitude 
according to the contribution of the $\sigma$ pole, Eq. (\ref{sig3}). 
The numerical results are shown in Fig. 5.

In order to verify whether the strong enhancement of the cross section 
entails an enhancement in the production rate of photon pairs per space-time 
element we calculate this quantity according to  Eq. (\ref{rate}) by folding
the cross section with the pion distribution function Eq. (\ref{rate}) and the 
relative velocity. The numerical results 
are shown in Fig. 6. The corresponding plot for the rates in the vacuum is 
not displayed, it is supressed by orders of magnitude.  This plot clearly 
indicates the strong enhancement of the photon production rate in a narrow 
temperature band around the Mott transition temperature. In our numerical 
example we chose $\mu = 80$ MeV where this effect happens at $T\approx 170$ 
MeV, see Fig. 6.  For temperatures below and above this critical value 
the resonance is broad and  not peaked. Hence, if we look to this process 
from the experimental point of view, the narrowing and enhancement of the 
$\sigma$ resonance along the transition line could be a hint to chiral 
symmetry restoration in heavy ion collisions.
 
\section{Conclusions}
\label{sec:sum}
In the present study we have investigated the medium dependence of the pion 
annihilation into a photon pair. 
Due to partial chiral symmetry restoration at finite temperatures 
and densities the $\sigma$ meson mass drops below the two pion threshold and 
appears as a bound state pole. This effect has several consequences. One of 
them is a strong enhancement of the cross section for pion annihilation in the 
two-photon channel as considered in the present work.

We have shown that the dramatic increase of the cross section for photon 
production appears in a narrow band in the $T-\mu$ plane which corresponds to 
the critical line of the chiral symmetry breaking/restoration transition. 
Thus the enhancement of photon pair production rate  at threshold should 
belong to the 
set of observable effects for the suspected chiral symmetry restoration in 
ultra-relativisitc heavy ion collision experiments. 
However, the  strong background due to $\pi^0\rightarrow 2\gamma$ complicates 
the observation of this effect in these experiments.

In order to check the relevance of the present studies for  experimental 
situations, a careful study of this process in a hydrodynamic scenario at the 
hadronisation transition is required. 
Such an analysis should include both the hadronic (pions) as well as quark 
effects under the influence of a hydrodynamical 
evolution of the hot and dense system and is deferred to future work.

\section{Acknowledgement}

M.K.V. is grateful to J. H\"ufner for fruitful discussions. One of us  (M.K.V.)
acknowledges  financial support provided by INTAS under Grant No. W 94-2915
and by the Max-Planck-Gesellschaft as well as the hospitality of the AG
''Theoretische Vielteilchenphysik'' at the University of Rostock where part
of this work has been done. E.A. Kuraev acknowledges financial support provided
by INTAS under Grant No. W 93-0239. S.S. is grateful for the financial support 
provided  by the MINERVA foundation.

%

\newpage
\vspace{2cm}

\begin{table}[hbt]
\begin{center}
\begin{tabular}{|c|c|c|c|c|c|c|}
\hline
Set \#&$T$ [MeV]&$ \mu$ [MeV]&$m$ [MeV]&$M_\pi$ [MeV]&$M_\sigma$ [MeV]
&$f_\pi$ [MeV]\\ \hline
1&0&0&300&140&610&93\\ \hline
2&100&252&154&165&338&55\\\hline
3&170&80&150&161&330&55\\\hline
\end{tabular}\\[1cm]
\caption{Typical examples for sets of parameter values 
obtained within the chiral quark model of Refs. \protect\cite{sbk,doc}.
Set \#1 corresponds to the vacuum case, the sets \#2 and \#3 correspond 
to points in the $T-\mu$ plane which are relevant for 
a discussion of the hadronization transition in heavy-ion collisions. }
\label{para1}
\end{center}
\end{table}

\newpage
\begin{figure}
\label{feyn1}
\centerline{\psfig{figure=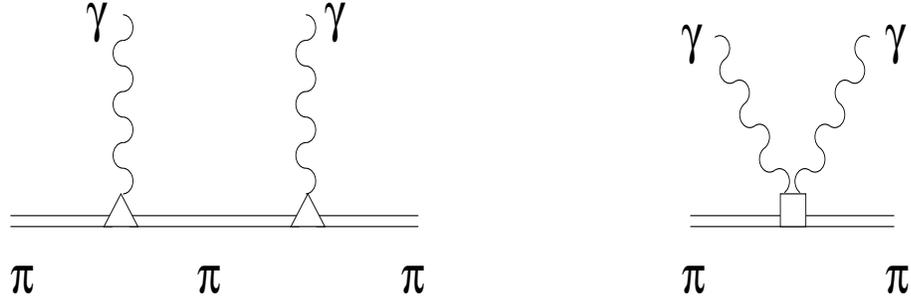,height=4cm,width=12cm,angle=-90}}
\caption[Fig. 1]{Feynman diagrams corresponding to the Lagrangian
(\protect\ref{L1}).}
\end{figure}
\begin{figure}
\label{feyn2}
\centerline{\psfig{figure=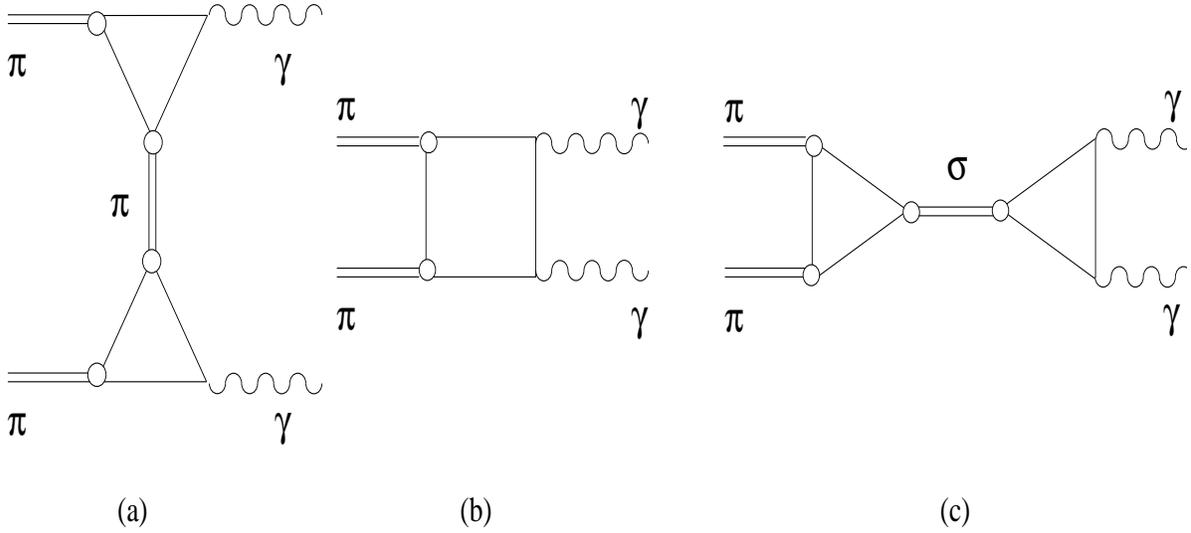,height=7cm,width=16cm,angle=-90}}
\caption[Fig. 2]{Feynman diagrams contributing to the process
$\pi \pi \to \gamma \gamma$ including quark substructure effects
in lowest order $(1/N_c)$ \protect{(ref{Lbox})-(\ref{L2})}. 
The lowest order gradient expansion of this set of diagrams corresponds 
to that of Fig. 1.}
\end{figure}
\begin{figure}[hbt]
\label{f7}
\centerline{\psfig{figure=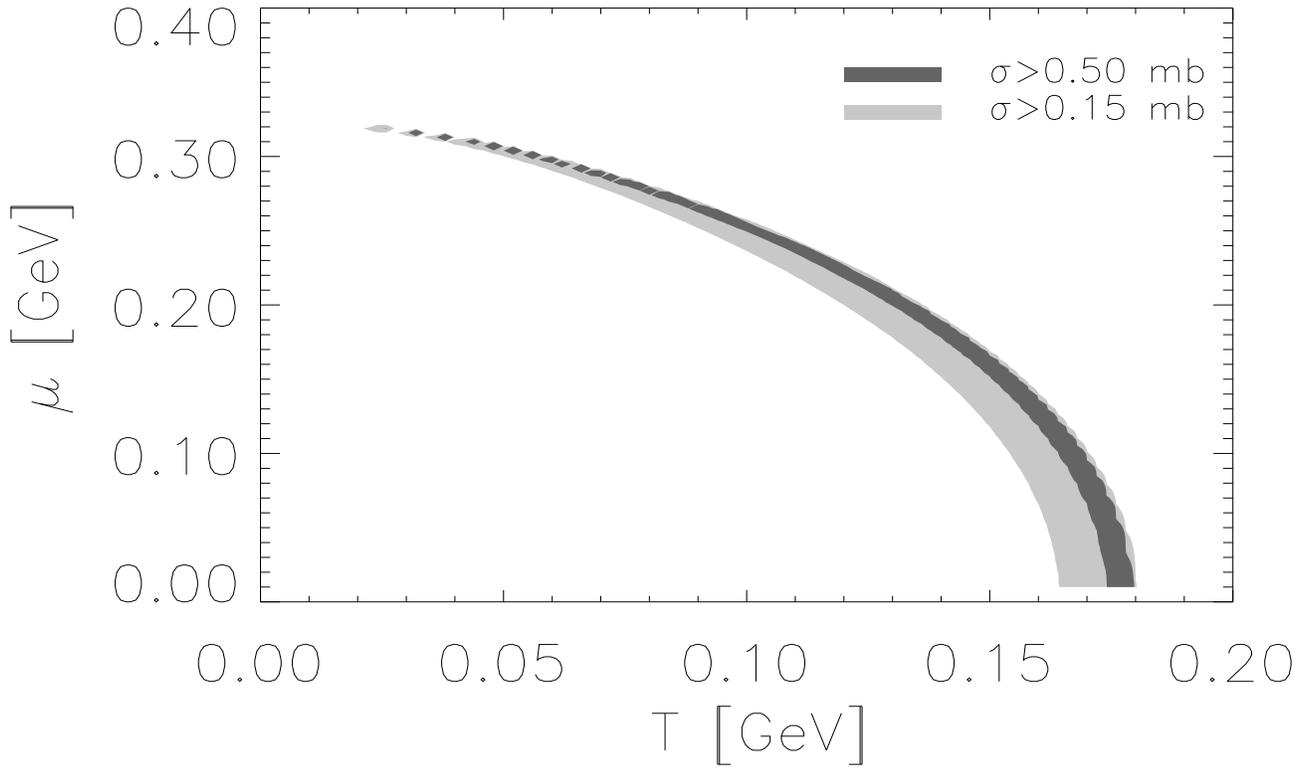,height=10.0cm,width=16cm,angle=0}}
\vspace{1cm}
\caption[Fig. 3]{Enhancement of the total cross section for 
$\pi\pi\to\gamma\gamma$ in the $T-\mu$ 
plane due to chiral symmetry restoration shown for 
$\sqrt{s - 4 M_\pi^2} = 10$ MeV .}
\end{figure}
\begin{figure}[hbt]
\label{cross}
\centerline{\psfig{figure=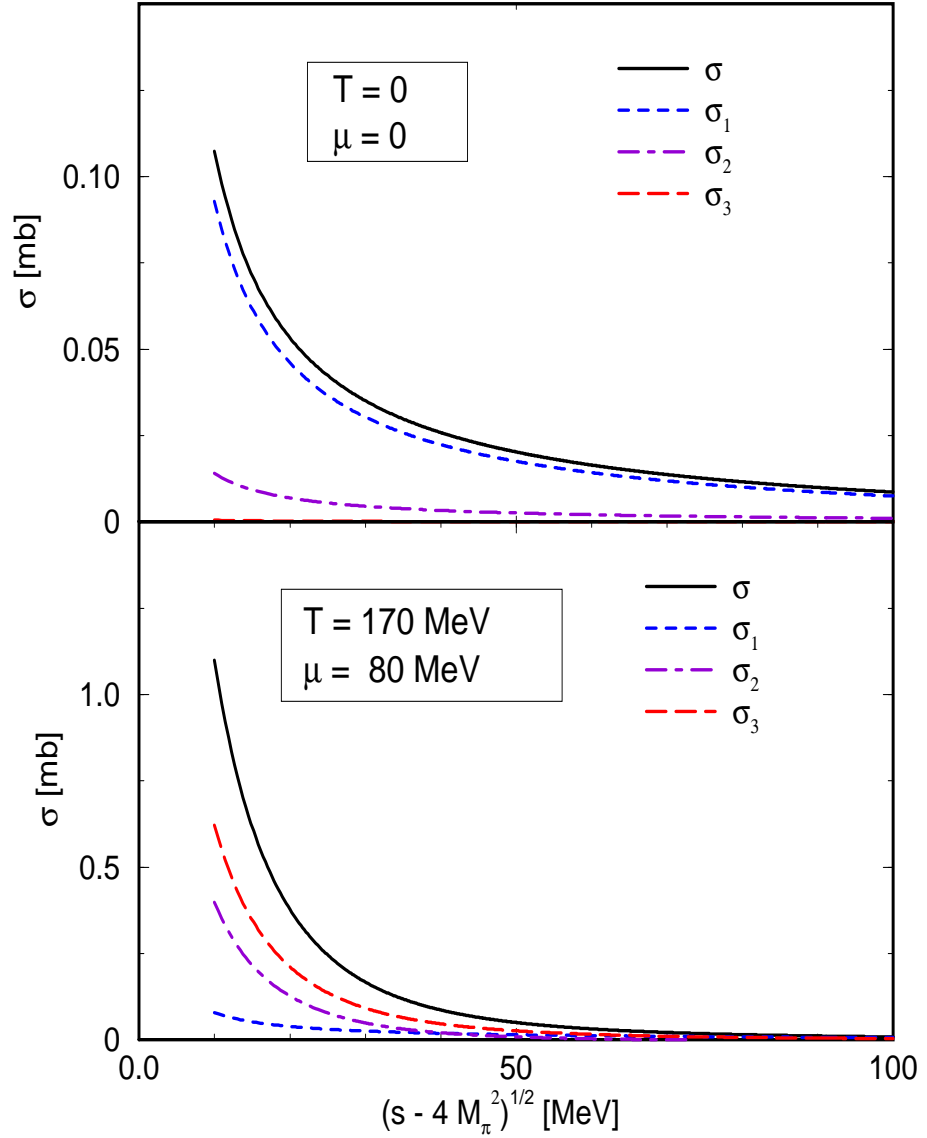,height=16.0cm,width=17cm,angle=-90}}
\vspace{1cm}
\caption[Fig. 4]{Cross section for the annihilation of charged pions into 
photon pairs in the vacuum $T, \mu= 0$ (upper panel) and in a hot, dense 
medium at $T=170$ MeV, $\mu=80$ MeV (lower panel).
This point in the $T-\mu$ plane 
has been chosen in accordance with recent analyses of hadron spectra at 
CERN-SPS \protect\cite{sps}.  The partial cross section for the Born process
($\sigma_1$) is dominant in the vacuum, the quark substructure contributions
contained in $\sigma_3$ become dominant at finite $T, \mu$ in the vicinity of 
the chiral restoration transition because of the strong enhancement of the 
$\sigma$ pole contribution. The interference contribution is denoted by 
$\sigma_2$.}
\end{figure}
\begin{figure}[hbt]
\label{cross2}
\centerline{\psfig{figure=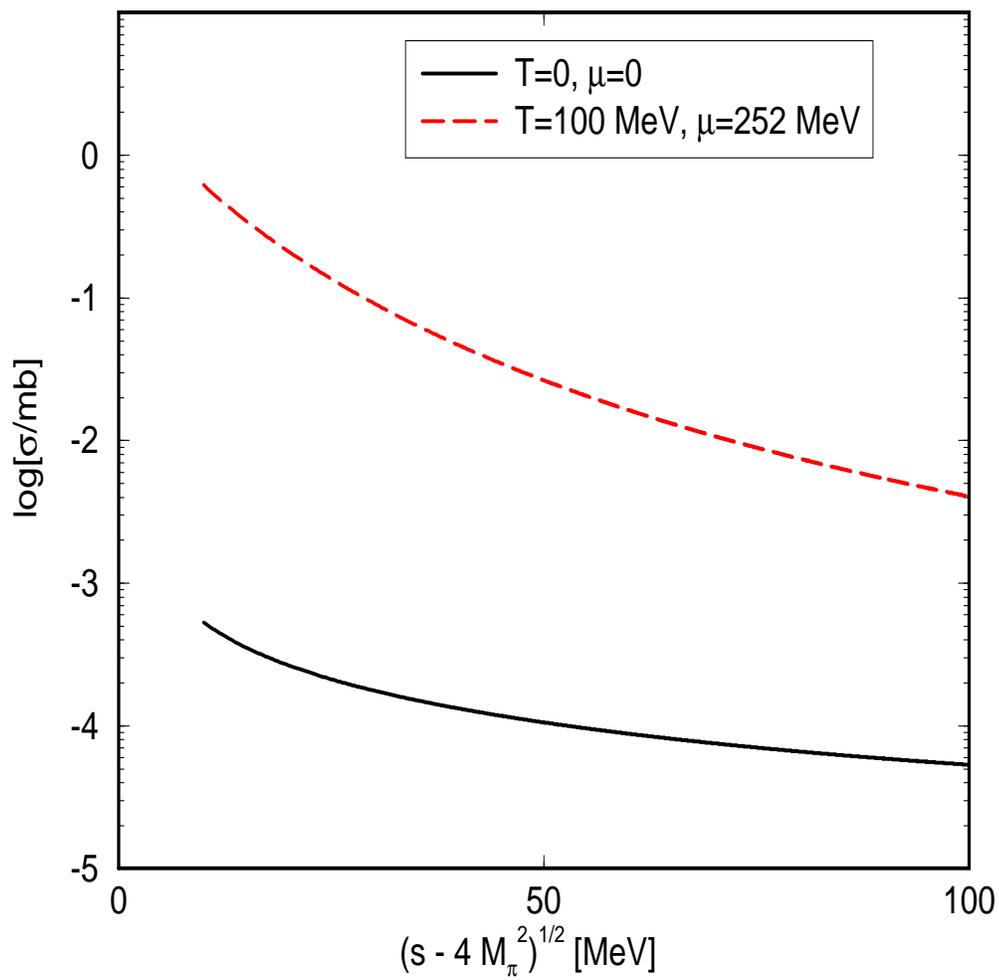,height=15.0cm,width=15cm,angle=0}}
\vspace{1cm}
\caption[Fig. 5]{Cross section for the annihilation of neutral pions into 
photon pairs in the vacuum $T, \mu= 0$  and in a hot, dense 
medium at $T=100$ MeV, $\mu=250$ MeV. }
\end{figure}
\begin{figure}[hbt]
\label{rates}
\centerline{\psfig{figure=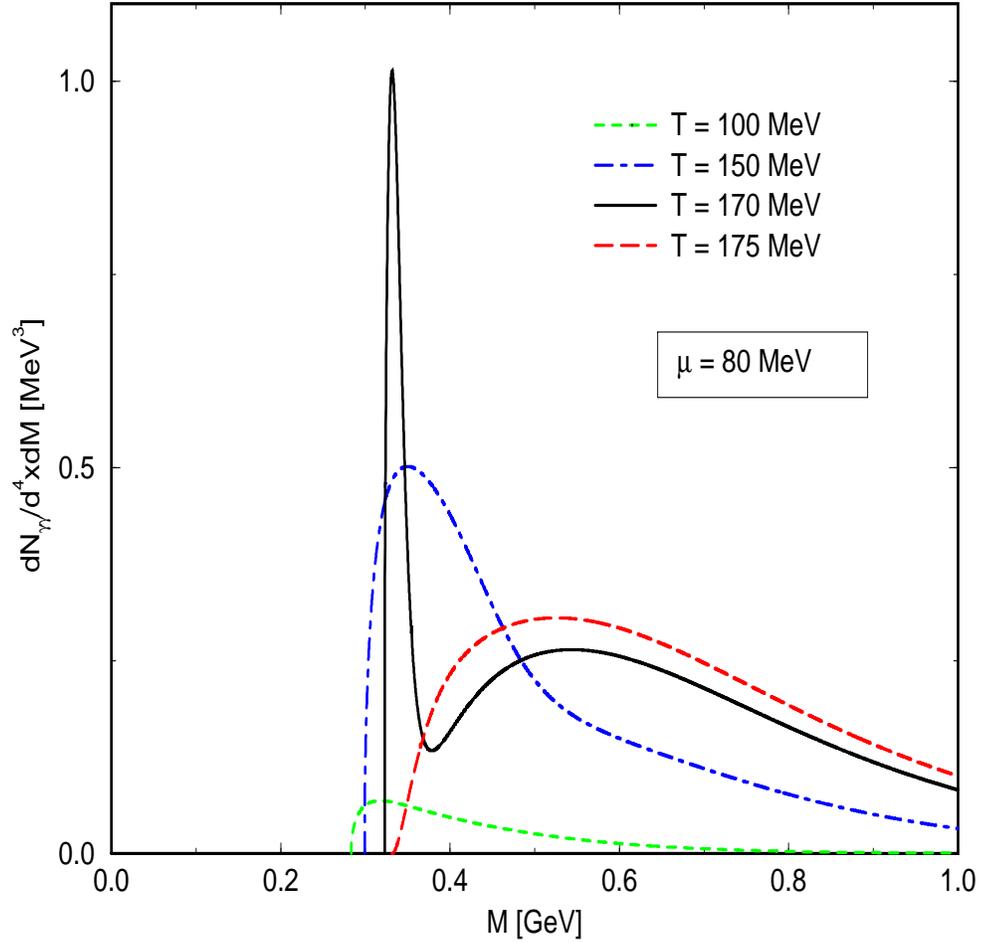,height=15.0cm,width=15cm,angle=-90}}
\vspace{1cm}
\caption[Fig. 6]{Production rate for photon pairs with invariant mass $M$ by
annihilation of charged pions for fixed chemical potential and different 
temperatures below and above the critical line, compare Fig. 3. }
\end{figure}

\end{document}